\documentclass[a4paper,11pt]{article}
\usepackage{jinstpub} 
\usepackage{lineno}

\title{Bayesian Optimization of The Relativistic Heavy Ion Collider Luminosity via $s^*$ Control }

\author[a]{Xiaofeng Gu\note{Corresponding author.}}
\author[a]{Benjamin Coe}
\author[a]{Guillaume Robert-Demolaize}
\author[a]{Takeshi Kanesue}
\author[a]{Masahiro Okamura}
\author[b]{William Fung}
\author[b]{Yue Hao}
\author[c]{Ji Qiang}
\author[c]{Xiaoye S. Li}
\author[c]{Yang Liu}

\affiliation[a]{Brookhaven National Laboratory,\\
Upton, NY 11973, USA}

\affiliation[b]{Department of Physics and Astronomy, Michigan State University,\\
East Lansing, MI 48824, USA}

\affiliation[c]{Lawrence Berkeley National Laboratory,\\
Berkeley, CA 94720, USA}

\emailAdd{xgu@bnl.gov}

\date{\today}
\abstract{
In order to aid in luminosity maximization at the interaction point (IP), the collision location $s_{IP}$ must be equivalent to the location of the minimum value of the beta function $s^*$. Accurate optics measurements and $s^*$ movements are therefore essential to luminosity optimization. However, according to current Relativistic Heavy Ion Collider (RHIC) operations measurements, average horizontal beta beat measurements between operating IPs are around $20\%$ along with significant variation in $s^*$ measurements. 
Precise control of the longitudinal beam waist position ($  s^*  $) has become a critical yet challenging knob for maximizing the luminosity of modern high-energy colliders, particularly with short bunch lengths and high crossing angles. We present the online application of Bayesian optimization (BO) using the GPTune framework to systematically tune sPHENIX luminosity via $  s^*  $ control at the Relativistic Heavy Ion Collider (RHIC). The GPTune framework was first validated at the RHIC Electron Beam Ion Source (EBIS), where it delivered up to a $70\%$ increase in beam intensity over the baseline, although experienced operators could also achieve this performance with longer manual tuning. The framework was then deployed during sPHENIX operations at RHIC. Using an intensity-normalized Zero-Degree-Calorimeter (ZDC) signal as the objective (owing to the unavailability of the live sPHENIX MVTX signal), GPTune successfully identified local luminosity maxima, recovered from intentionally degraded $  s^*  $ configurations, and uncovered residual horizontal and vertical waist offsets in the interaction region. These results demonstrate the robustness and efficiency of Bayesian optimization for real-time collider tuning under noisy and time-varying conditions. The developed methodology with $  s^*  $ control provides a powerful new tool for precision luminosity optimization and will be especially valuable for next-generation short-bunch colliders such as the Electron-Ion Collider (EIC).}


\begin{document}
\maketitle
\flushbottom
\section{\label{sec: Introduction} Motivation}

\begin{figure}[htbp]
    \centering
    \includegraphics*[width=1\columnwidth]{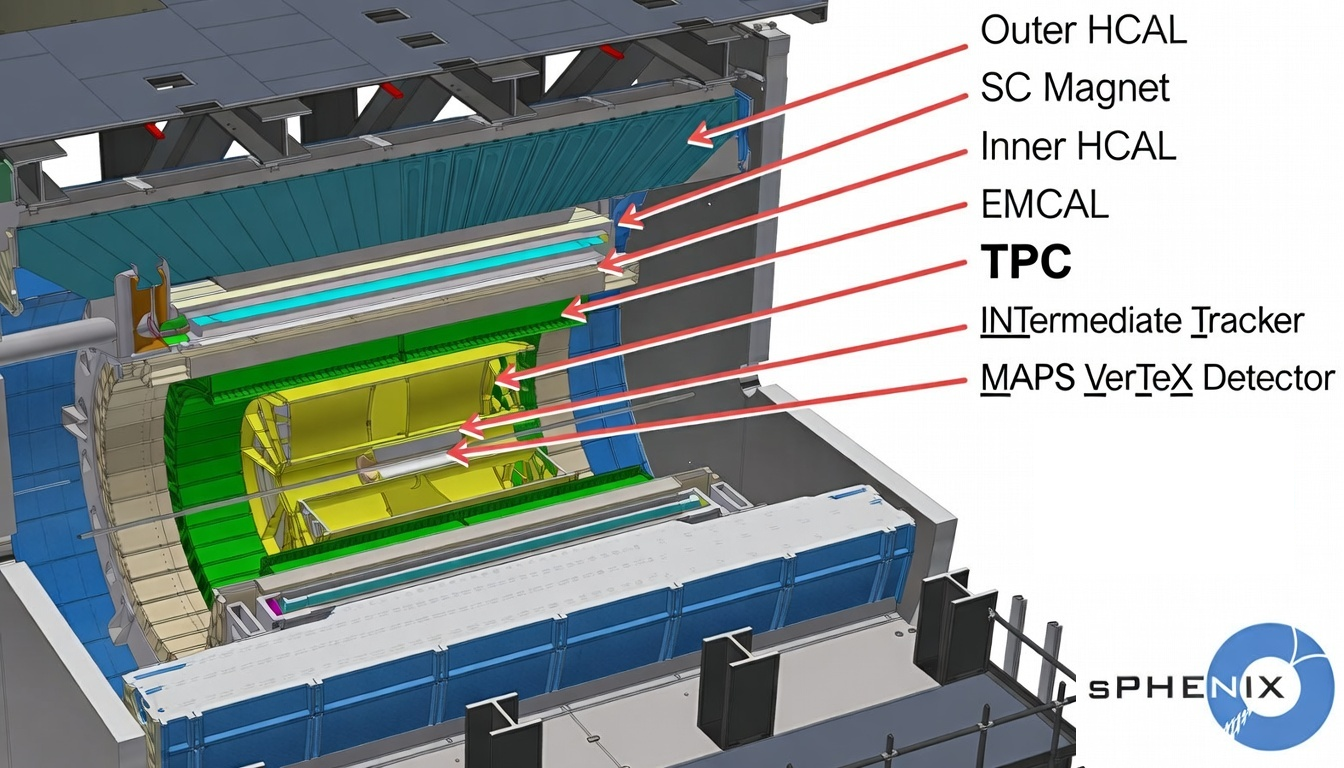}
    \caption{\label{fig:pictures/sPHENIX_Device} Overview of the three sPHENIX tracking systems. The MVTX is the innermost detector, with an acceptance within $|s| < 0.1\,\mathrm{m}$ of the interaction point (IP). Courtesy of D. Morrison.}
\end{figure}

 A state-of-the-art jet detector named sPHENIX ~\cite{NSAC2015,Kim2019, sPHENIX} was proposed, commissioned, and operated at the Relativistic Heavy Ion Collider (RHIC) from 2023 to 2025. This detector featured precision tracking and calorimetry that enable high-statistics studies of the Quark-Gluon Plasma through jet modification, upsilon suppression, and open heavy flavor production. The innermost component of the three sPHENIX tracking systems is the Monolithic-Active-Pixel-Sensor-based Vertex Detector (MVTX) (Fig.~\ref{fig:pictures/sPHENIX_Device}), which has an acceptance within $|s| < 0.1\,\mathrm{ m}$ of the interaction point (IP)~\cite{Klest2020}.

A collider's performance is fundamentally determined by the bunch intensity, the beam sizes (or $\beta^*$) at the collision point, the bunch length, and the longitudinal waist position ($s^*$) within the interaction region (IR) \cite{HerrMuratori1996}. Previous studies on $s^*$ location control indicated that the uncorrected $s^*_y$ could be as high as $0.29\text{ m}$ in the RHIC Yellow ring \cite{Hao2015}. While this was not a critical limitation for past RHIC operations, the shorter bunch length enabled by the 56~MHz RF cavity increases the sensitivity to $s^*$ offsets. For the current $|s^*| < 0.1\text{ m}$ requirement, a relative offset of $0.5\text{ m}$ between the two beams in either the Blue or Yellow ring could reduce the luminosity recorded by the MVTX detector by approximately $10\%$ due to reduced acceptance. Consequently, precise control of the beta-waist location ($s^*$) has been introduced as a novel optimization knob for the sPHENIX MVTX luminosity.

In RHIC operation, luminosity optimization is a multidimensional process involving the adjustment of established machine and beam parameters. While these parameters have been routinely optimized during machine operation, the longitudinal waist position \(s^*\) has not previously been used as an independent optimization knob at RHIC. Waist-shift studies have nevertheless been reported at other colliders, including the LHC~\cite{CoelloDePortugal2020MD3287} and SuperKEKB~\cite{Zhou2024LuminositySuperKEKB}. Building on these studies, the present work investigates the controlled adjustment of \(s^*\) as an additional degree of freedom for luminosity optimization at RHIC. We develop a systematic procedure for determining the optimum waist position and evaluate its impact on luminosity under realistic RHIC operating conditions.

Meanwhile, Bayesian optimization (BO) methods have been widely adopted within the particle accelerator community to address the challenges of tuning complex, high-dimensional systems with noisy, expensive-to-evaluate objectives~\cite{ Duris2020, Shalloo2020, Hanuka2021, Roussel2021, Gao2022, Miskovich2022, Huang2021, Xopt, Roussel2024BOAccel, Xu2023BOInjection, Valenta2025LWFA_BO, Wang2026RFBayesianCavity}. These applications demonstrate BO's versatility across linacs, synchrotrons, FELs, and rare isotope facilities, yielding remarkable success in improving beam quality at many facilities worldwide.

Therefore, to rapidly optimize RHIC performance by controlling $s^*$, GPTune~\cite{GPTune2025, Liu2021}, a Bayesian optimization framework developed at LBNL, was implemented to address this multi-objective optimization problem. Bayesian optimization was selected over traditional gradient-based methods due to its superior sample efficiency in noisy experimental environments and its capability to navigate non-convex objective landscapes without becoming trapped in local sub-optima. Furthermore, Bayesian optimization operates effectively from cold-start conditions with zero prior historical data, initializing its surrogate model entirely online through initial random evaluations.

In this paper, we present the implementation and evaluation of online sPHENIX luminosity optimization at RHIC using GPTune via $s^*$ control. The paper is structured according to the experimental procedure. In Section~\ref{sec: Optics tuning}, the GPTune algorithm is introduced and validated through beam intensity optimization at the RHIC Electron Beam Ion Source (EBIS)~\cite{Alessi2010} in Section~\ref{sec: Validation GPTune}. Section~\ref{sec: Experimental Results} details the implementation of the GPTune algorithm for RHIC sPHENIX Zero-Degree-Calorimeter (ZDC) signal optimization. Finally, Section~\ref{sec: Discussion} provides a brief discussion and concluding remarks.

\section{\label{sec: Optics tuning}  The GPTune Framework}
 
Given a complex $d$-dimensional objective function $y(\mathbf{x})$, Bayesian optimization (BO) models the objective using a Gaussian process (GP) prior:
\begin{equation}
f(\mathbf{x}) \sim \mathcal{GP}\left(\mu(\mathbf{x}), \Sigma(\mathbf{x},\mathbf{x}')\right).
\end{equation}
In its simplest form, we assume a zero mean, $\mu(\mathbf{x})=0$, and use a radial basis function (RBF) covariance kernel:
\begin{equation}
\Sigma(\mathbf{x},\mathbf{x}') =
\sigma^{2}\exp\left(
-\sum_{i=1}^{d}\frac{(x_i-x'_i)^2}{2l_i^2}
\right),
\end{equation}
where $\sigma^2$ characterizes the overall variation of the objective function and $l_i$ is the characteristic length scale associated with the $i$-th optimization parameter. A small $l_i$ indicates that the objective function can vary rapidly with $x_i$, whereas a large $l_i$ indicates a smoother dependence on that parameter.

Given $N$ observed function evaluations $\mathbf{X}=[\mathbf{x}_1,\mathbf{x}_2,\ldots,\mathbf{x}_N]^T$ and $\mathbf{y}=[y_1,y_2,\ldots,y_N]^T$, conditioning the GP prior on these observations produces a posterior GP. Its predictive mean at an unsampled point $\mathbf{x}^*$ is
\begin{equation}
\mu^* =
\Sigma(\mathbf{x}^*,\mathbf{X})
\Sigma(\mathbf{X},\mathbf{X})^{-1}\mathbf{y},
\end{equation}
with a corresponding predictive variance that quantifies the model uncertainty at $\mathbf{x}^*$. In practice, the GP hyperparameters, including the length scales $l_i$ and the overall scale $\sigma^2$, can be optimized using methods such as Maximum Likelihood Estimation (MLE). Optimizing these hyperparameters improves the GP representation of the observed objective-function data and can therefore affect the efficiency of the subsequent optimization.

The posterior GP provides the predicted objective value and its uncertainty to an acquisition function, which determines the next sampling point. The acquisition function balances exploitation of regions with favorable predicted objective values and exploration of regions with greater model uncertainty. The choice of acquisition function and GP hyperparameters can therefore influence the efficiency and convergence of the optimization. After a new function evaluation is obtained, it is added to the observed data, and the posterior model is updated iteratively.

In our experiments, $y(\mathbf{x})$ represents either the EBIS linac intensity signal ($d=9\text{--}19$ power-supply parameters) or the sPHENIX luminosity signal ($d=2$ control parameters).

GPTune is a Python-based Bayesian optimization autotuning framework designed for expensive black-box functions arising in high-performance computing (HPC) codes, machine learning frameworks, and control experiments. GPTune was originally developed to tune software libraries that contain numerous tuning parameters and require substantial resources for each function evaluation. It was later extended to accelerator simulations \cite{Kan24, Luo2026, liu2023detecting} and experimental control \cite{Gu2024IPAC}. In addition to supporting a variety of covariance kernels and optimizers, GPTune includes advanced features such as multi-task learning \cite{Liu2021}, transfer learning \cite{Liu2021, cho2023harnessing}, multi-objective tuning, multi-fidelity tuning \cite{Zhu2023}, input uncertainty modeling \cite{Luo2026}, ensembles of multiple models \cite{cho2023harnessing}, additive GPs \cite{luo2024non}, hybrid models \cite{luo2024hybrid}, and a shared historical database \cite{cho2021enhancing}.

Utilizing a Python interface developed at RHIC, GPTune communicates with the RHIC control system to directly manipulate hardware parameters. In this paper, GPTune was operated in Lite mode with Latin-hypercube sampling using the SampleLHSMDU sampler ~\cite{GPTune2025, Liu2021}. A Gaussian Process (GP) surrogate model with a radial basis function (RBF) kernel was used to model the objective function. Fixed random seeds were used for both the sampling and modeling stages to ensure reproducibility. The optimization was performed using GPTune's multi-level allocation (MLA) framework, which selected successive parameter configurations based on the GP surrogate and the Expected Improvement (EI) acquisition function. The EI acquisition function was optimized using the particle swarm optimization (PSO) search algorithm provided by the SearchPyGMO search class.

\section{\label{sec: Validation GPTune} Validation of GPTune at the EBIS}
The EBIS is a compact and versatile ion accelerator that serves as the pre-injector system for both RHIC and the NASA Space Radiation Laboratory (NSRL). To evaluate and validate GPTune prior to its implementation for luminosity optimization during sPHENIX operations, the algorithm was deployed at the EBIS in 2023.

\begin{figure*}[!tbh]
    \centering
    \includegraphics*[width=\textwidth]{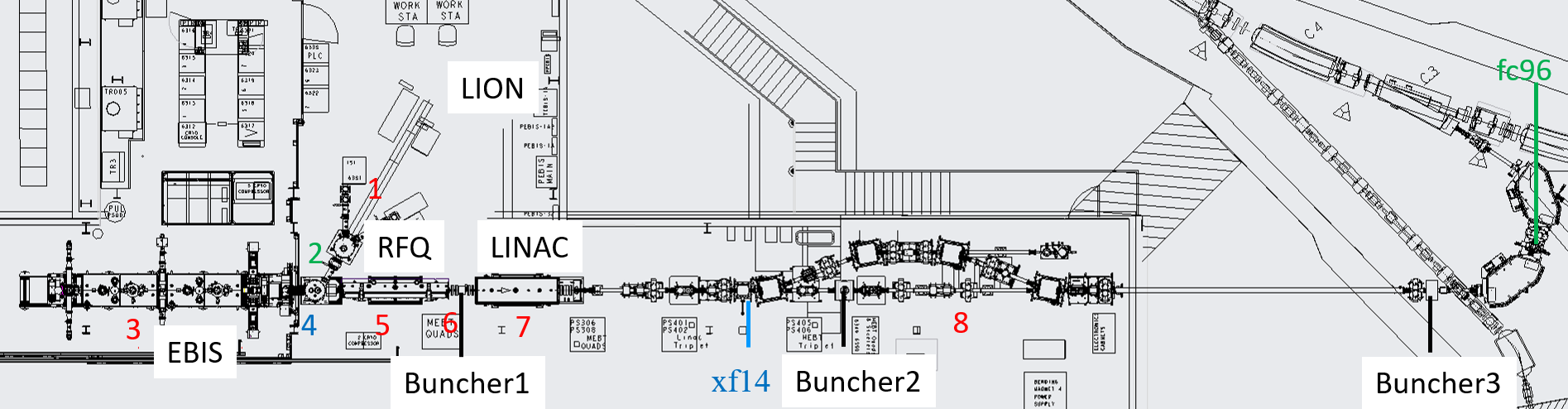}
    \caption{Layout of the EBIS facility. The numbers indicate the section numbers. Section 1 is the Laser Ion Source, which provides the ion beam. Section 2 denotes the injection beamline upstream of the EBIS source (Section 3), which steers the beam into the EBIS. Section 4 is the extraction beamline, which transports the extracted beam to the RFQ. Sections 5 and 6 correspond to the RFQ and Buncher 1, respectively, while Sections 7 and 8 represent the LINAC and transport line, respectively.}
    \label{fig:EBIS_Layout}
\end{figure*}

Fig.~\ref{fig:EBIS_Layout} illustrates the layout of the EBIS facility. The pre-injector system comprises several beamline sections, including the Laser Ion Source (LION), the injection line, the EBIS source, the extraction line, a Radio Frequency Quadrupole (RFQ), the Medium Energy Beam Transport (MEBT), and the High Energy Beam Transport (HEBT). A current transformer (xf14) and a Faraday cup (fc96) were utilized to measure the ion beam intensity during the optimization process. The different diagnostics were selected based on their availability during the respective optimization measurements.

During the optimization process, a $^{28}\text{Si}^{11+}$ ion beam was used.The Booster--AGS beam injection system has a supercycle period of 7~s. Some power supplies require two supercycles (14~s) for their outputs to stabilize after a parameter change. Once the power supplies have stabilized, the script acquires four measurements for averaging, with each measurement separated by one supercycle plus an additional 1~s (8~s). Thus, each GPTune iteration requires approximately 46~s for power-supply stabilization and beam-intensity measurements.

To establish a baseline and reduce the dimensionality of the problem, the injection and extraction beamlines were initially optimized separately. Table~\ref{tab:Control_Parameters} lists the control parameters used for GPTune optimization of the EBIS injection (fc96) and extraction (xf14) beamlines. Although some parameter names are common to both beamlines, they refer to corresponding control elements that are physically distinct in the injection and extraction lines. Nine variables are optimized for the injection line and ten for the extraction line.

\begin{table}[ht]
    \centering
    \caption{Parameters for GPTune Optimization}
    \label{tab:Control_Parameters}
    \small
    \begin{tabular}{lcc}
        \hline
        \textbf{Parameter} & \textbf{Injection [fc96]} & \textbf{Extraction [xf14]} \\
        \hline
        IonLens20-40kV & \checkmark & \checkmark \\
        DeflPlatBias & \checkmark & \checkmark \\
        16PoleX & \checkmark & \checkmark \\
        16PoleY & \checkmark & \checkmark \\
        Gridded\_Lens & \checkmark & \checkmark \\
        Horiz\_Bend\_Defl & \checkmark & \checkmark \\
        Inter\_Vert\_Defl & \checkmark & \checkmark \\
        Inter\_Vert\_Defl\_Lower & \checkmark & \checkmark \\
        Horiz\_Sphere\_Bend & \checkmark & $\times$ \\
        RFQ\_Horiz\_Bend & $\times$ & \checkmark \\
        LEBT\_Solenoid & $\times$ & \checkmark \\
        \hline
        \textbf{Total Variables} & \textbf{9} & \textbf{10} \\
        \hline
    \end{tabular}
\end{table}

For the EBIS optimization, the objective function \(y(\mathbf{x})\) is the measured beam intensity at the corresponding diagnostic location: \(I_{\mathrm{fc96}}\) for injection-line optimization and \(I_{\mathrm{xf14}}\) for extraction-line optimization. GPTune varies the corresponding machine-control parameters to maximize the measured intensity.

\subsection{Independent Optimization of the Injection and Extraction Lines}

\begin{figure}[!tbh]
    \centering
    \includegraphics*[width=\columnwidth]{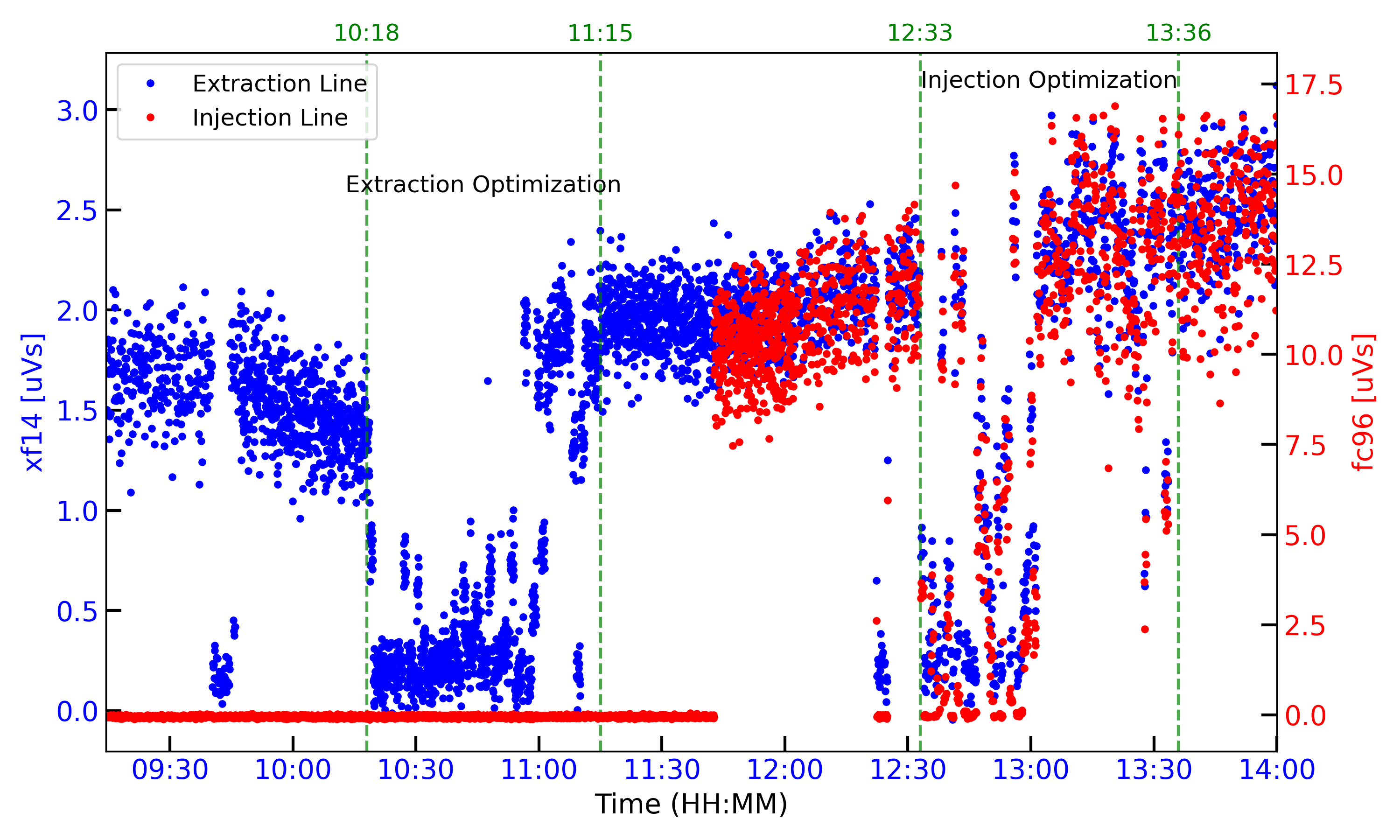}
    \caption{Optimization progress for the EBIS injection and extraction lines. Blue dots indicate the current transformer (xf14) signals during extraction line tuning, while red dots represent the Faraday cup (fc96) signals during injection line tuning. The specific intensity measurement device for each run was selected based on its availability during the respective optimization measurements. }
    \label{fig:EBIS_Injection_Extraction_Optimization}
\end{figure}

The blue dots in Fig.~\ref{fig:EBIS_Injection_Extraction_Optimization} illustrate the xf14 beam intensity signal during the extraction line optimization process (the interval between the first two vertical green dashed lines). This process involved $d=10$ control parameters and was conducted over 70 iterations. The average beam intensity exhibited a substantial increase from 1.4 $\mu$Vs to 2.0 $\mu$Vs, representing a remarkable $43\%$ improvement.

Following the successful optimization of the extraction line, we proceeded to optimize the injection line while holding the optimized extraction line power supply settings constant. 

The red dots in Fig.\ref{fig:EBIS_Injection_Extraction_Optimization} display the beam intensity signal during the injection line optimization process (the interval between the third and fourth vertical green dashed lines). For the injection line optimization, the Faraday-cup (fc96) measurement was employed as the objective function, using $d=9$ control parameters over 60 total iterations. The average beam intensity increased from 11.5 $\mu$Vs to 14.0 $\mu$Vs (as shown on the right vertical axis), representing an improvement of approximately $22\%$.

It is important to note that while xf14 and fc96 share the same units ($\mu$Vs), a direct comparison of their absolute values is not feasible because different normalization factors were applied during data processing.

Meanwhile, as shown in Fig.~\ref{fig:EBIS_Injection_Extraction_Optimization}, the beam-intensity signal exhibits temporal fluctuations during the optimization. A downward trend is observed during one optimization period (the blue dots), whereas an upward trend occurs during another (the red dots), indicating that the temporal variation is not a simple monotonic drift. Such fluctuations are therefore treated as part of the stochastic variation of the experimental objective.

\begin{figure}[!tbh]
    \centering
    \includegraphics*[width=\columnwidth]{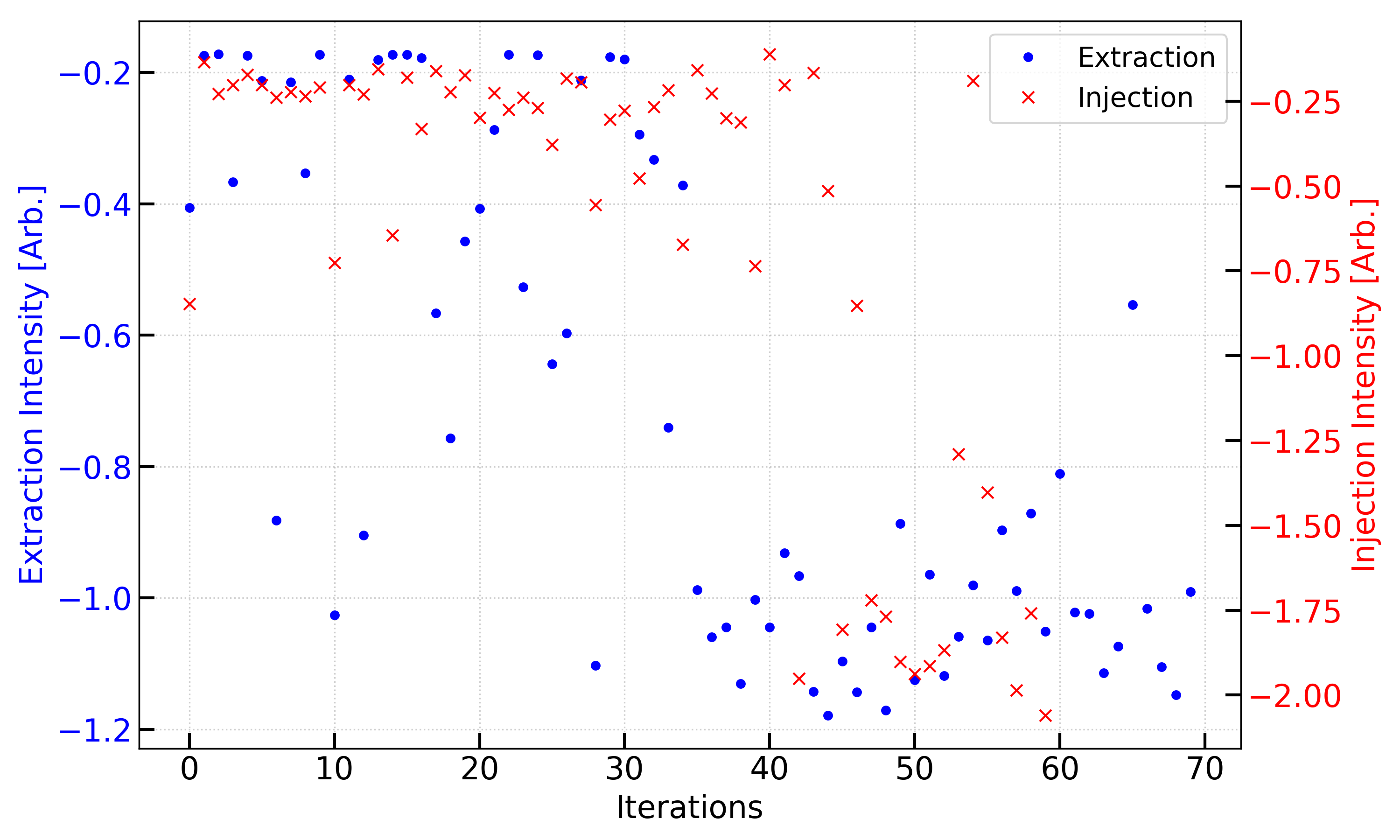}
    \caption{GPTune iteration progresses during the extraction- and injection-line optimizations. The evolution of the objective function and selected dominant tuning parameters demonstrates the progression of the optimization toward a stable operating region. Increasingly negative values of the normalized intensity signal correspond to higher beam intensity.}
    \label{fig:GPTune_Iterations_Converger}
\end{figure}

Fig.~\ref{fig:GPTune_Iterations_Converger} shows the relative intensity improvement for the extraction and injection lines as a function of the GPTune optimization iterations. The horizontal axis represents the iteration number, while the vertical axis displays the normalized intensity signal. Because GPTune is configured to minimize the objective function to achieve maximum output, a more negative value indicates a better optimization result. 

Because the influence of external accelerator conditions could not be eliminated during the measurements, to evaluate the effects of the optimized power-supply settings on beam intensity, the optimized settings from the injection and extraction line optimizations were applied in three machine configurations. The original default settings represent the machine state before optimization, the extraction line only (Ext.) configuration uses the optimized extraction-line parameters while retaining the default injection-line settings, and the Inj + Ext configuration combines the optimized injection- and extraction-line parameters; The Inj + Ext configuration was subsequently reverted to the original settings and remeasured to assess the reproducibility of the observed beam intensity.

\begin{figure}[!ht]
    \centering
    \includegraphics[width=\columnwidth]{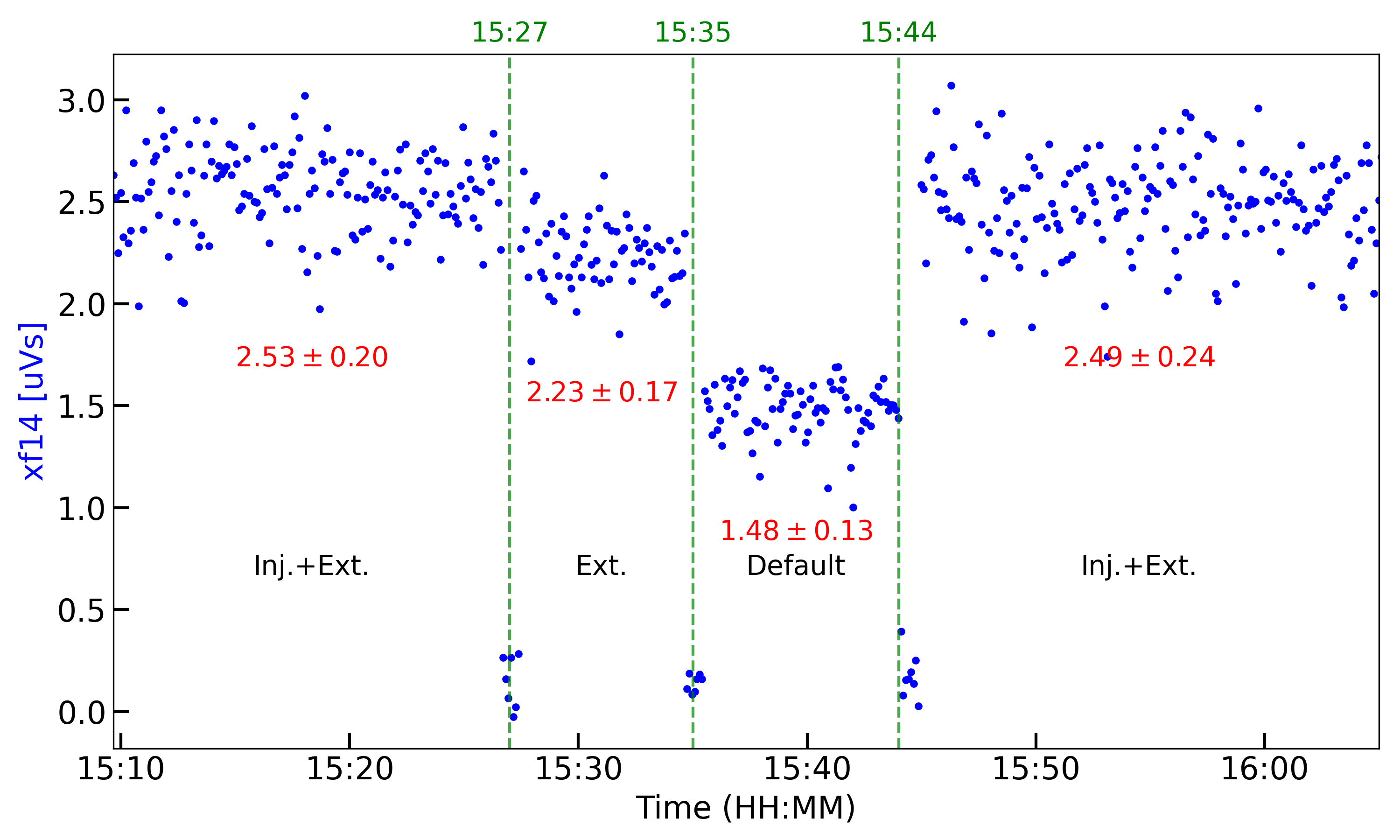}
        \caption{Comparison of beam intensity across three configurations: combined injection and extraction line optimization (Inj + Ext), extraction line only (Ext), and the original default (Default) settings. The red values indicate the mean beam intensity and the respective standard deviations for each machine state. To demonstrate reproducibility, the Inj + Ext configuration was reverted to the original settings and the beam intensity was remeasured.}
    \label{fig:Injection_Extraction_Result}
\end{figure}

The results for these three configurations are presented in Fig.~\ref{fig:Injection_Extraction_Result}. As observed in the figure, significant intensity gains were recorded by the xf14 measurement, which increased from a baseline of 1.48 $\mu$Vs to 2.49 $\mu$Vs or 2.53 $\mu$Vs following the combined optimization of both beamlines. This represents a substantial $68-71\%$ intensity improvement over the initial default configuration.
    
Meanwhile, Fig.~\ref{fig:EBIS_Injection_Extraction_Optimization} and Fig.~\ref{fig:Injection_Extraction_Result} reveal a characteristic of the beam intensity signal: substantial noise. The signal exhibits a standard deviation of $10\%$ and a peak-to-peak deviation of $15\%$. The Gaussian Process (GP) model underlying GPTune accounts for uncertainty in the objective measurements, providing an advantage for optimization in the presence of stochastic experimental fluctuations.

\subsection{Simultaneous Optimization of Injection and Extraction Lines}

Following the independent optimization of the injection and extraction beamlines, wherein the optimized settings from the previous stages were maintained as the new baseline, we proceeded to optimize both sections simultaneously by tuning all $d=19$ power supplies at once. This global optimization approach was designed to account for any interdependencies between these two sections and to evaluate GPTune's performance in a higher-dimensional parameter space. 

The simultaneous optimization was initially configured for 80 iterations. However, to ensure full convergence within the high-dimensional parameter space, an additional 10 iterations were appended to the process, resulting in a total of 90 iterations. This flexibility is a advantage of the GPTune framework, as it allows for real-time adjustments based on the observed progress.

\begin{figure}[!ht]
\centering
\includegraphics[width=\columnwidth]{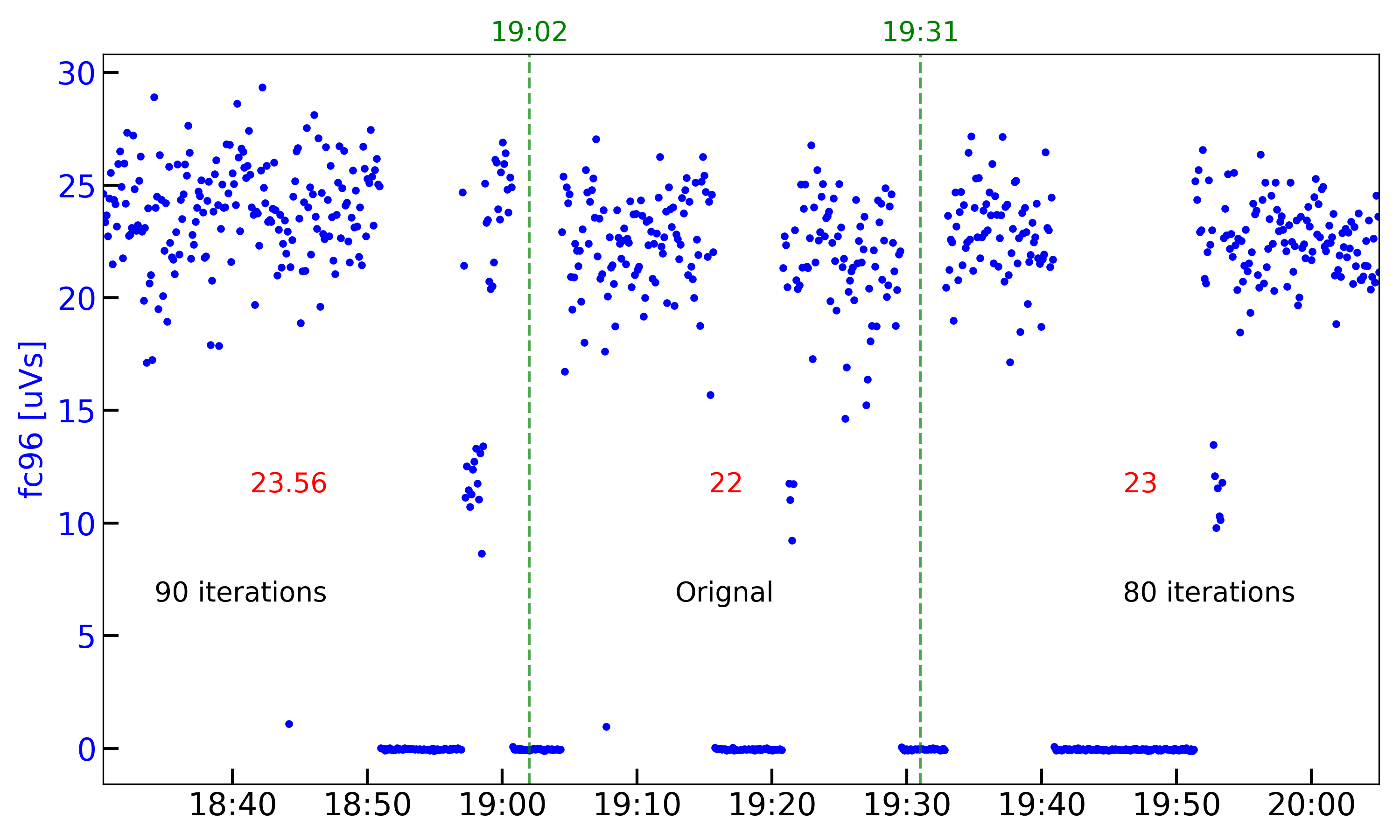}
\caption{Optimization results achieved through the simultaneous tuning of $d=19$ control parameters across both the injection and extraction lines. The red values indicate the mean beam intensity for three distinct machine configurations: the original baseline, the intermediate 80-iteration state, and the final 90-iteration optimized state.}
\label{fig:Injection_Extraction_Combined_Result}
\end{figure}

Fig.~\ref{fig:Injection_Extraction_Combined_Result} illustrates the results of the simultaneous optimization of both the injection and extraction lines ($d=19$ control variables) using the fc96 signal as the objective function. Starting from the previously optimized state, the initial average signal was approximately 22 $\mu$Vs. After 80 iterations, the averaged signal increased to 23 $\mu$Vs, and after 90 iterations, it further improved to 23.56 $\mu$Vs. 

Thus, the reported \(7.1\%\) (from 22 $\mu$Vs to 23.56 $\mu$Vs) improvement represents a further improvement over the previously optimized configuration obtained through two separate injection- and extraction-line optimization, rather than the total improvement from the original machine settings.

There are two primary reasons why the simultaneous optimization did not yield another dramatic jump in intensity. First, both the injection and extraction lines had already been independently optimized in the previous stages, meaning the system was already operating near a local optimum. Second, the optimization target was defined by the \texttt{fc96} signal, which is located significantly downstream from these beamlines; consequently, downstream bottlenecks can constrain the overall efficiency of the optimization. 

\subsection{Section Summary}

In this section, the GPTune framework was applied to EBIS intensity optimization, achieving a $70\%$ improvement in beam intensity at the xf14 current transformer (CT)~\cite{Gu2024IPAC}. This success demonstrated the effectiveness of GPTune as a robust optimization tool for accelerator operations, validating its suitability for the subsequent sPHENIX luminosity optimization tasks.

It is important to emphasize that while an experienced EBIS operator can achieve comparable performance through manual tuning over the course of one to two weeks, each optimization stage conducted by GPTune required only a few hours of beam time. This demonstrates that GPTune is a valuable tool for sPHENIX luminosity optimization, where manual tuning is both time-consuming and costly.

\section{\label{sec: Experimental Results} Luminosity Optimization in A Collider}

\subsection{Control of $s^*$ Using RHIC Interaction-Region Quadrupoles}

\begin{figure}[!ht]
    \centering
    \includegraphics[width=\columnwidth]{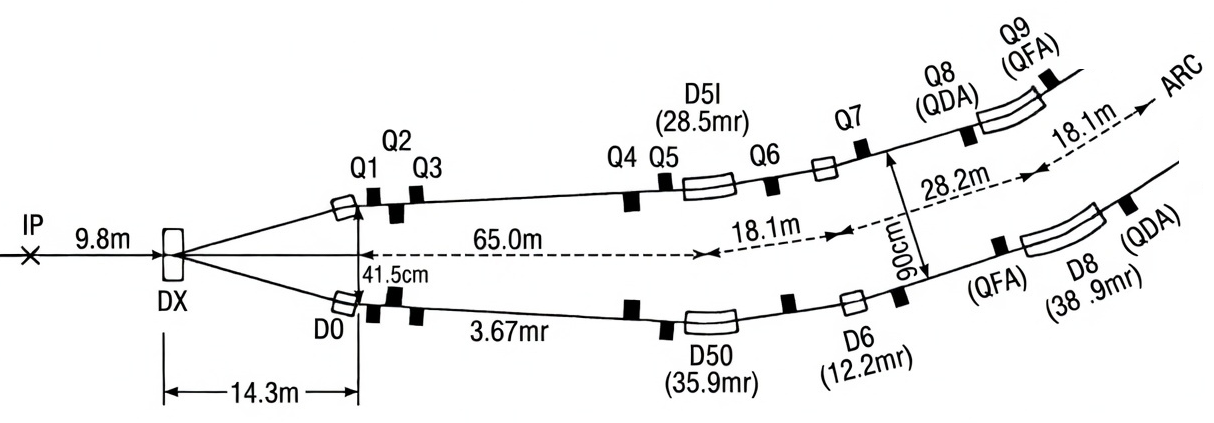}
    \caption{Layout of one half of a RHIC interaction region, showing the
    quadrupole magnets used to control the longitudinal waist position $s^*$.}
    \label{fig:RHIC_IR_Layout}
\end{figure}

To control the longitudinal waist position \(s^*\), some selected magnets in the RHIC interaction region (IR) are used. Figure~\ref{fig:RHIC_IR_Layout} shows the layout of one half of the RHIC interaction region. Each half-insertion consists of a dispersion-matching section containing Q6--Q8, a betatron-function matching section containing Q1--Q5, and the beam-crossing dipoles D0 and DX ~\cite{RHICConfigManual}. 

In the present study, the Q1--Q8 quadrupoles on both sides of the interaction point are used to adjust the transverse optics and thereby control the horizontal and vertical waist positions, \(s_x^*\) and \(s_y^*\). The framework presented here utilizes an optics response matrix to perform systematic and controlled $s^*$ adjustments~\cite{fung2026}. The optics response matrix provides the relationship between quadrupole-strength changes and the resulting changes in \(s_x^*\), \(s_y^*\), and other constrained optics parameters. By solving this response relation for a desired waist displacement, appropriate changes in the quadrupole power-supply currents can be determined and applied to the IR magnets. This approach enables \(s^*\) to be shifted systematically while minimizing unwanted perturbations to the surrounding optics~\cite{fung2026}.

The GPTune optimization loop used in this study is described as follows. GPTune provides the requested \(s^*\) change, which is converted by the optics response matrix into the corresponding quadrupole power-supply changes. After the settings are applied to the RHIC IR magnets, the luminosity signal is measured and used to calculate the objective function. This measurement is returned to GPTune, which updates its model and proposes the next \(s^*\) setting, forming a closed-loop optimization process.Two control variables ($d=2$) optimized by GPTune in this study are the horizontal and vertical $s^*$ of the Blue ring in RHIC.

\subsection{Optimization Challenges and Hardware Limitations}

The application of Bayesian optimization to the sPHENIX luminosity optimization was subject to critical hardware constraints and dynamic beam dynamics that set it apart from the controlled EBIS environment.

First, the optimization of luminosity faced significant hardware constraints. A several-minute delay existed between the RHIC control system and the local MVTX signal processing system—a latency issue that could be resolved in future runs, rendering the live MVTX signal unavailable for real-time optimization. Furthermore, the 56 MHz cavity was not operational during the experiment, which limited the ability to shorten the bunch length.

Consequently, the sPHENIX ZDC rate was employed as the single objective function, prioritizing luminosity maximization. The Zero-Degree-Calorimeter (ZDC) is located downstream of the RHIC interaction point along the beam direction and detects neutral particles produced at very small angles relative to the beam axis. The ZDC signal provides a proxy for the collision rate and is therefore correlated with the instantaneous luminosity.

Second, the transition from the controlled EBIS environment to the dynamic storage ring environment introduced another primary technical challenge: a non-stationary target signal. Unlike the steady-state intensity at EBIS, the luminosity signal at sPHENIX is a "drifting" target that naturally decays over time due to the depletion of beam intensity and the growth of both transverse and longitudinal emittance.

To ensure a model-independent objective function, a custom normalized luminosity was adopted for the subsequent 2025 experimental runs. The optimization target was changed to an intensity-normalized luminosity, defined as:
\begin{equation}
\label{Eq.Normalized_Lumi}
L_{\text{norm1}} = \frac{\text{ZDC}}{I_{\text{Blue}} \cdot I_{\text{Yellow}}}
\end{equation}

\subsection{Luminosity Optimization}

\subsubsection{$s^*$ Measurement and Luminosity Optimization under Induced Initial Offsets}

To further evaluate the effectiveness and robustness of the GPTune framework, an experiment was conducted during the 2025 RHIC run using 100 GeV gold (Au) beams. The intensity-normalized luminosity, as defined in Eq.~(\ref{Eq.Normalized_Lumi}), was used as the optimization target.

Prior to the GPTune optimization, it was necessary to confirm that $s^*$ shifted as requested. Since optics measurements at RHIC can only be performed without active beam-beam collisions, a dedicated $s^*$ calibration phase was conducted at the beginning of the experiment.

\begin{figure}[!ht]
\centering
\includegraphics[width=\columnwidth]{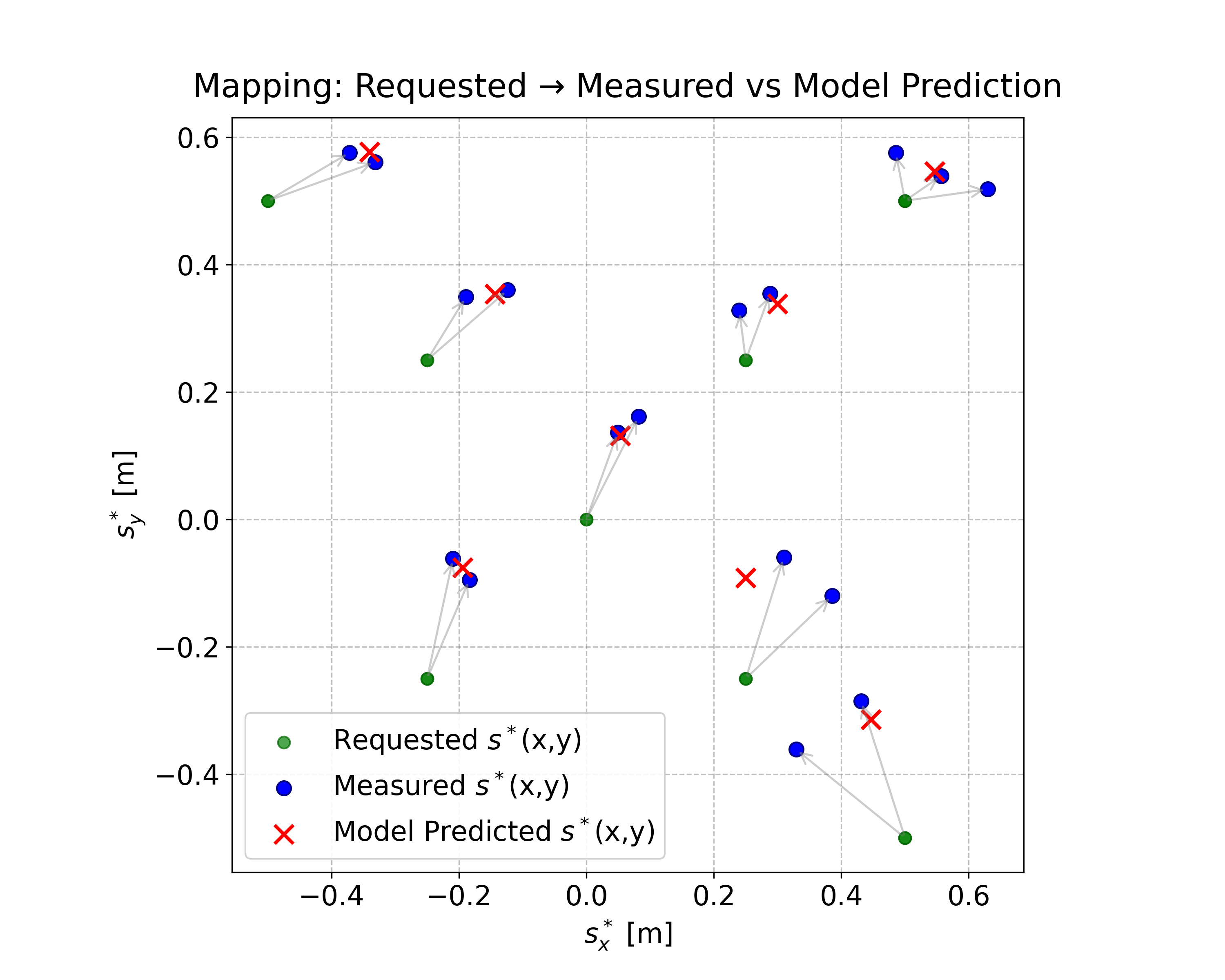}
\caption{Validation of the $s^*$ control model: The nominal requested $s^*$ setpoints (green dots) are compared against the actual $s^*$ values extracted from optics measurements (blue dots). The regression model (red crosses) demonstrates a robust reconstruction of the physical $s^*$ position.}
\label{fig:s_Star_Fitting}
\end{figure}

As illustrated in Fig.~\ref{fig:s_Star_Fitting}, 16 measurements of \(s^*\) (blue dots) were obtained for 8 discrete requested \(s^*\) values (green dots), with two measurements taken at each requested value. A model-independent Ordinary Least Squares (OLS) linear regression method was used to calculate the $s^*$ values from turn-by-turn data~\cite{fung2026}, which was subsequently cross-validated using the beta-from-amplitude method. The OLS method determines the best-fit model coefficients by minimizing the sum of squared differences between the measured beam turn-by-turn positions and the predicted linear response.

To bridge the gap between "requested" and "actual" positions during active collisions, when direct optics measurements are unavailable during the GPTune optimization process, these calibration results were modeled and predicted (red crosses) using a nominal linear regression framework.

As seen in Fig.~\ref{fig:s_Star_Fitting}, systematic offsets are observed between the measured and requested \(s^*\) positions, possibly due to magnet hysteresis and/or other machine-related errors. These discrepancies, ranging approximately from 0.1 m to 0.2 m in both the horizontal and vertical planes, are consistent with previous optics studies of the Yellow ring~\cite{Hao2015}. This implies that the luminosity can be further optimized by tuning $s^*$ away from the default RHIC configuration, where the requested $s_{xy}^*$ is currently set to $(0\text{ m}, 0\text{ m})$~\cite{RHICConfigManual,fung2026, madxmanual}.

\begin{figure}[!ht]
\centering
\includegraphics[width=\columnwidth]{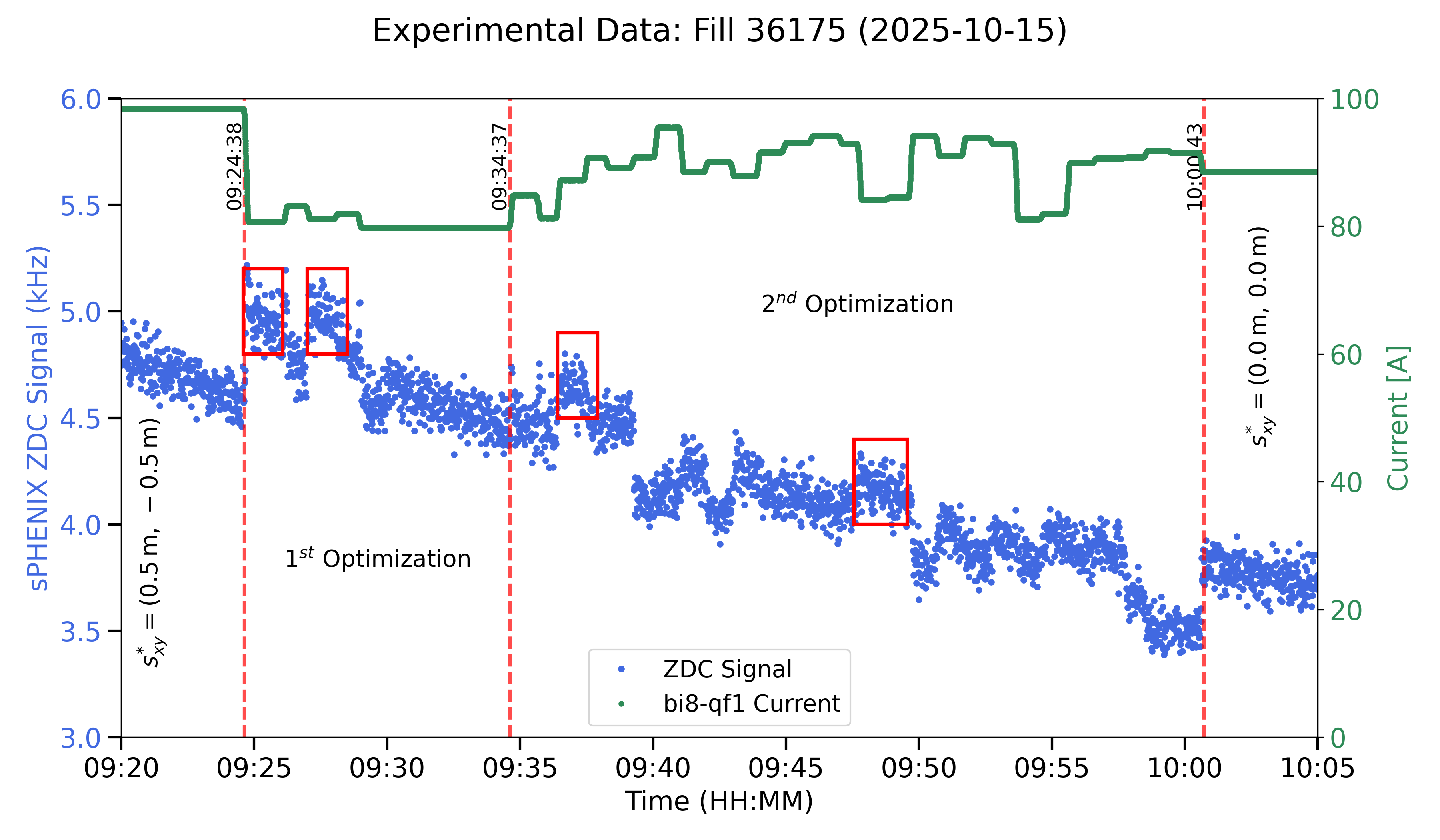}
\caption{Operational data during the sPHENIX $s^*$ waist optimization (Fill 36175). The plot shows the sPHENIX ZDC coincidence rate objective function (blue dots) alongside a representative interaction region power supply current (green line) indicating $s^*$ knob adjustments. Red highlighted boxes indicate the active GPTune optimization phases where parameter tuning yields localized luminosity peaks.}
\label{fig:APEX_36175_Data}
\end{figure}

Following the optics measurements, the search efficiency and convergence of the GPTune framework were evaluated by intentionally offsetting the Blue beam $s^*$ to a luminosity degraded starting position of $s_{xy}^* = (0.5\text{ m}, -0.5\text{ m})$, the final position from the previous $s^*$ calibration measurement. As illustrated in Fig.~\ref{fig:APEX_36175_Data}, two successive optimization cycles were executed; subsequently, the parameters were manually restored to their original values of $s_{xy}^* = (0\text{ m}, 0\text{ m})$ to provide a comparison.

Fig.~\ref{fig:APEX_36175_Data} shows the iteration history of the sPHENIX zero-degree calorimeter (ZDC) coincidence rate (blue dots) and a representative interaction region (IR) power supply current (green line), which is displayed as a primary indicator of the $s^*$ waist shift. 

In the first iteration of the first optimization interval (the first red box), the ZDC signal recovered to a level above the intentionally degraded baseline, demonstrating the ability of GPTune to recover luminosity from a degraded starting condition. In the following iterations, the repeated identification (highlighted by the red boxes) of favorable operating points demonstrates that GPTune was able to identify and converge toward a local optimum within a limited number of iterations under evolving machine conditions.

Although achieving substantially higher luminosity is inherently constrained by hardware and machine limitations, the ability to efficiently reach a high-performing operating point within a limited number of iterations demonstrates the practical effectiveness of the GPTune framework.

\subsubsection{Luminosity Optimization with Baseline Resets for Comparative Analysis}

During the third phase of the experiment, historical models were intentionally disabled, forcing the GPTune framework to perform an autonomous exploration of the parameter space. 

\begin{figure}[!ht]
\centering
\includegraphics[width=\columnwidth]{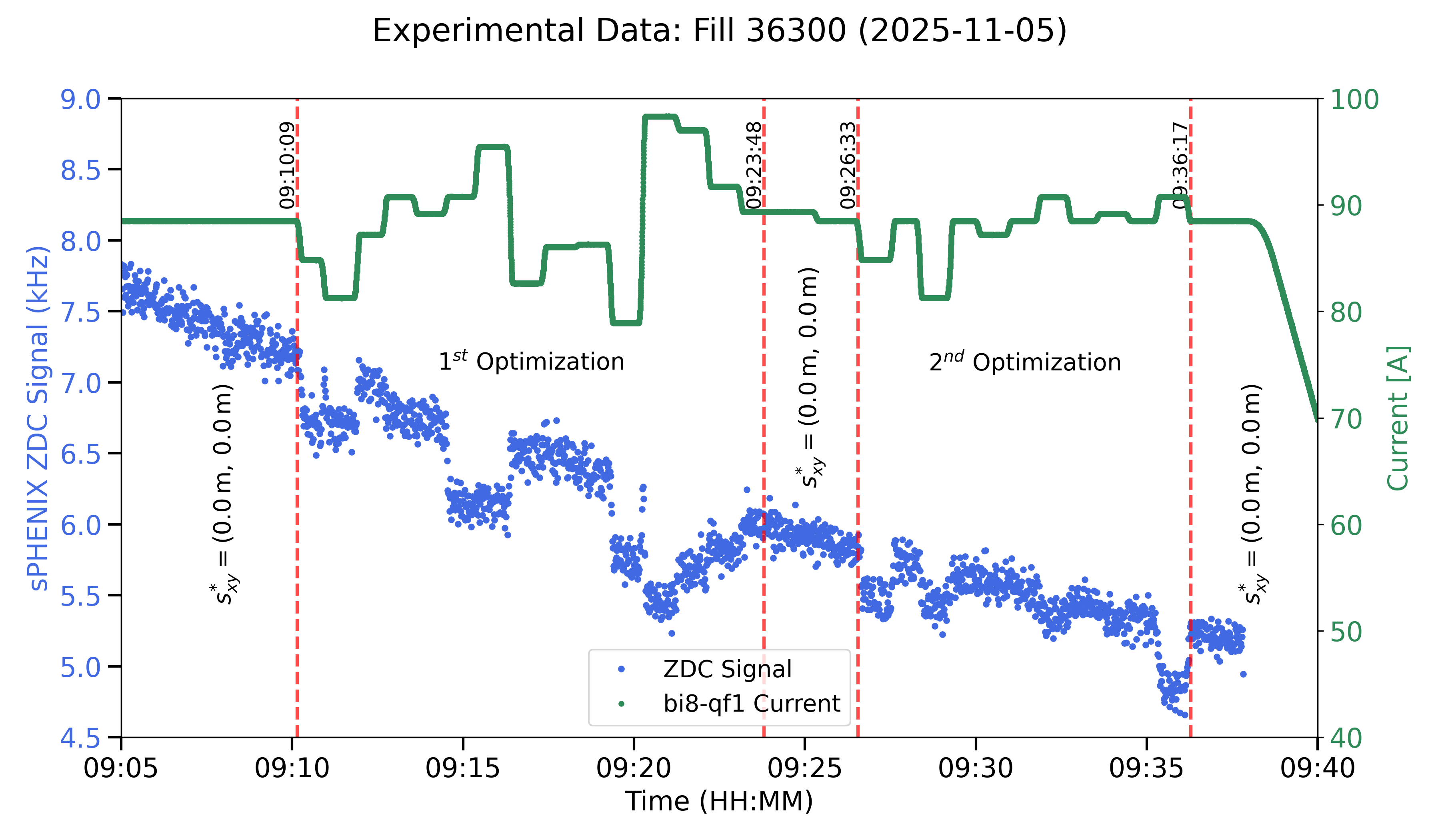}
\caption{Operational data for the baseline validation run (Fill 36300). The plot correlates the sPHENIX ZDC signal with the current of a representative IR power supply, providing a direct comparison between machine settings and luminosity optimization progress.}
\label{fig:APEX_36300_Data}
\end{figure}

\begin{figure*}[!ht]
\centering
\includegraphics*[width=\textwidth]{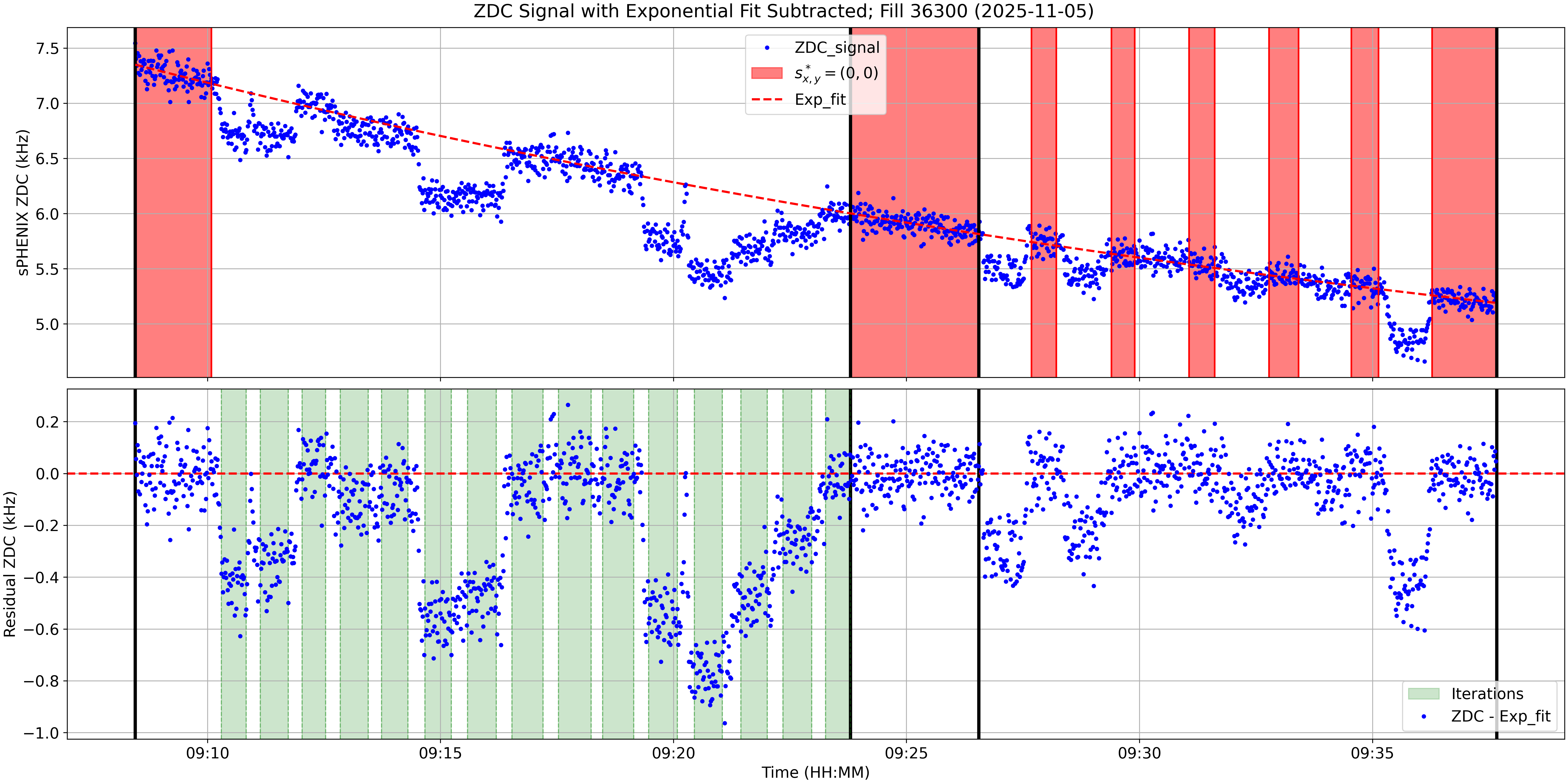}
\caption{The top plot displays the sPHENIX ZDC signal during the optimization procedure; the red areas indicate periods where $s_{x,y}^*$ was reset to the nominal $(0\text{ m}, 0\text{ m})$ baseline. The red dashed line represents the exponential fit of these baseline points ($R^2 = 0.987$), serving as a time-dependent reference for the natural beam decay. The bottom plot illustrates the residual ZDC signal after subtracting this decay model. The green shaded regions in the first optimization cycle denote the 15 GPTune data acquisition windows—each approximately 30 seconds in duration—conducted once the IR8 power supplies had fully settled.}
\label{fig:GPTune_Optimization_sPHENIX}
\end{figure*}

To quantify the optimization gains relative to the machine's default configuration, a series of baseline resets were integrated into the experimental timeline. As illustrated in Fig.~\ref{fig:APEX_36300_Data}, $s^*$ was returned to the nominal setting of $s^*_{x,y} = (0\text{ m}, 0\text{ m})$ at the beginning, middle, and end of the experiment. Furthermore, during the second optimization cycle, the baseline reference of  $s^*_{x,y} = (0\text{ m}, 0\text{ m})$ was implemented every 60 seconds (indicated by the red areas in the top plot of Fig.~\ref{fig:GPTune_Optimization_sPHENIX}).

To isolate the optimization gains from the natural beam decay, all ZDC data points corresponding to the nominal \(s_{xy}^*=(0\,\mathrm{m},0\,\mathrm{m})\) baseline were fitted with an exponential decay function plus a constant baseline,
\begin{equation}
ZDC_{\mathrm{fit}}(t)=A e^{-t/\tau}+C,
\end{equation}
where \(A\) is the amplitude of the decaying component, \(\tau\) is the decay time constant, and \(C\) is the constant baseline. The fit achieved an \(R^2\) value of 0.987. As shown by the red dashed line in the top plot of Fig.~\ref{fig:GPTune_Optimization_sPHENIX}, this fit provides a time-dependent reference for the machine's baseline performance throughout the store.

The bottom plot of Fig.~\ref{fig:GPTune_Optimization_sPHENIX} shows the residual sPHENIX ZDC signal after subtracting the fitted exponential decay component from the original raw ZDC signal. The residual is defined as
\begin{equation}
R(t) = S_{\mathrm{ZDC}}(t)-S_{\mathrm{fit}}(t),
\end{equation}
where \(S_{\mathrm{ZDC}}(t)\) is the original raw ZDC signal and \(S_{\mathrm{fit}}(t)\) is the exponential decay model fitted to the signal. The residual \(R(t)\) therefore represents the variation in the ZDC signal that is not accounted for by the underlying exponential decay and is used to characterize the luminosity-related changes during the optimization.

A positive residual ZDC signal indicates that the measured luminosity is higher than that expected from the nominal exponential decay model, whereas a negative residual indicates lower luminosity. As shown in the bottom plot of Fig.~\ref{fig:GPTune_Optimization_sPHENIX}, GPTune identified favorable operating points with residual ZDC signals near or above zero. These points were found in close proximity to the machine's default configuration of $s^*_{x,y}=(0\text{ m},0\text{ m})$, indicating that the default waist position is already near a favorable region of the luminosity response.

This convergence was achieved reliably across both the first and second optimization cycles; however, as shown in the residual plot, achieving further absolute luminosity gains beyond this nominal peak proved challenging right now.

The residual ZDC rate can be plotted as a function of the requested $s^*$ values to examine the  ependence of the luminosity signal on the longitudinal waist position. This provides a direct visualization of the ZDC response to changes in $s^*$ and allows the optimization results to be evaluated in terms of the waist-position dependence.

\begin{figure}[!ht]
\centering
\includegraphics[width=\columnwidth]{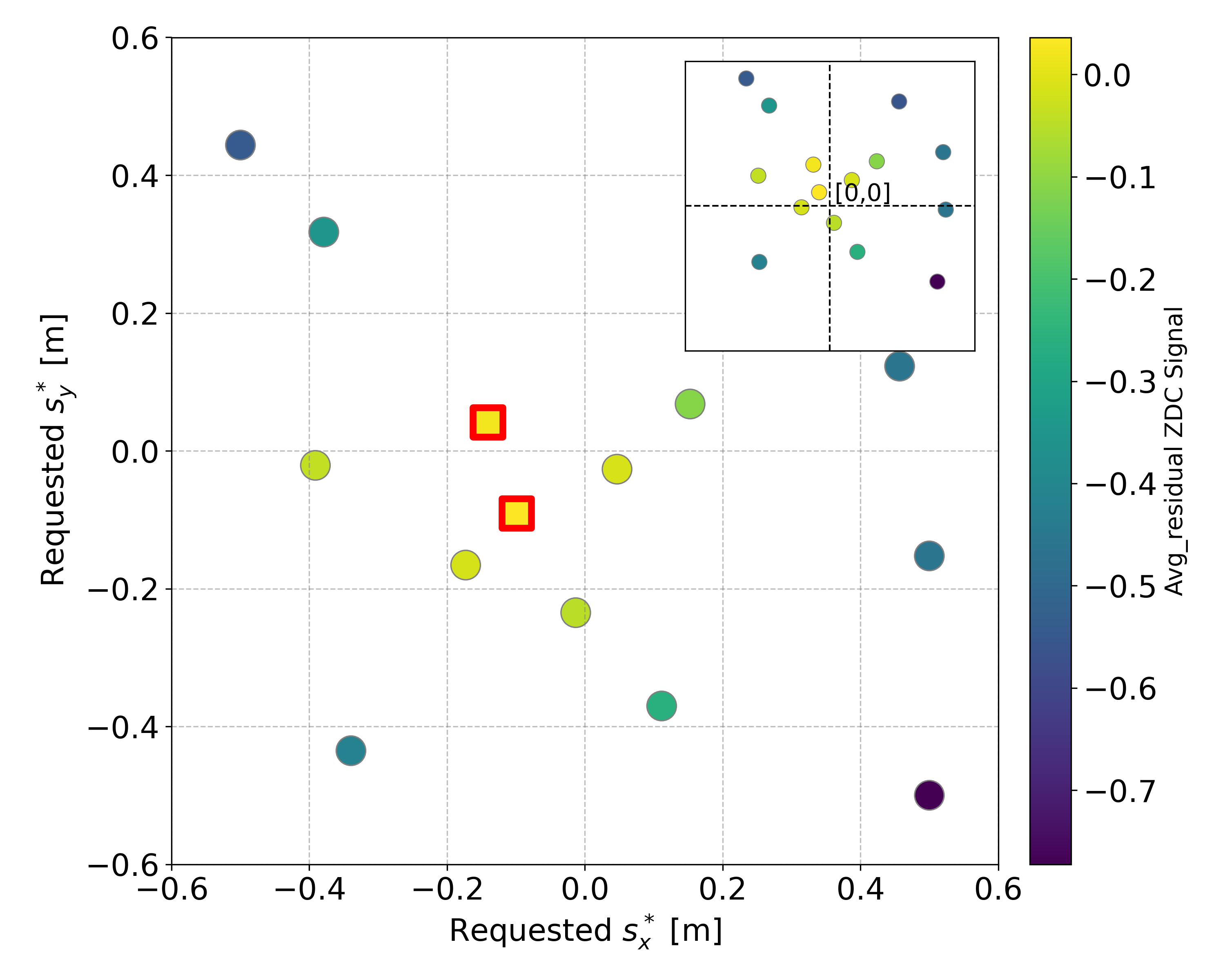}
\caption{Normalized ZDC signal rate as a function of the requested $s_{xy}^*$ during optimization, confirming again the presence of offsets in both planes of $s^*$ at the location of maximum ZDC rate. The top-right plot shows the correlation between the averaged residual ZDC rate and the predicted $s_{xy}^*$ obtained from the linear regression model developed in the previous section.}
\label{fig:s_Star_ZDC}
\end{figure}

The 15 averaged residual ZDC values were obtained from 15 individual GPTune iterations, each corresponding to one of the 15 green-shaded regions in Fig.~\ref{fig:GPTune_Optimization_sPHENIX}. For each iteration, the requested $s_{x,y}^*$ setting was applied, and the residual ZDC signal was averaged over the $30\text{--}40$~s interval following stabilization of the IR8 power supplies. Thus, each green-shaded region represents one GPTune iteration and one objective-function evaluation. The resulting dependence of the averaged residual ZDC signal on the GPTune-requested $s_{x,y}^*$ values is shown in Fig.~\ref{fig:s_Star_ZDC}.

The statistical noise level was determined from the standard deviation of the residual ZDC values obtained from the 15 green-shaded regions, yielding \(0.078~\mathrm{kHz}\). Two configurations yielded positive averaged residual ZDC values, \(0.036\pm0.078~\mathrm{kHz}\) and \(0.021\pm0.078~\mathrm{kHz}\), as indicated by the two red boxes in Fig.~\ref{fig:s_Star_ZDC}. Although these values provide preliminary indications of improved luminosity, their magnitudes are smaller than the statistical standard deviation and therefore remain within the statistical noise of the measurement.

Meanwhile, the observation that the maximum ZDC rate occurs at a non-zero requested $s^*$ setpoint provides a preliminary indication of residual $s^*$ offsets in the IR8 region. These findings, while requiring further statistical validation, are consistent with the analysis presented in Fig.~\ref{fig:s_Star_Fitting}, which suggests that the actual luminosity peak may be shifted from the nominal $s^*_{x,y} =(0\text{ m}, 0\text{ m})$~\cite{RHICConfigManual,madxmanual}.  

By applying the regression model derived in Fig.~\ref{fig:s_Star_Fitting} to the $s_{x,y}^*$ values requested during the GPTune optimization, the predicted $s_{x,y}^*$ coordinates were reconstructed and correlated with the averaged residual ZDC, as illustrated in the top-right panel of Fig.~\ref{fig:s_Star_ZDC} which range is also $\pm0.6m$ for both horizontal and vertical axis.

Therefore, Fig.~\ref{fig:s_Star_ZDC} qualitatively indicates that the GPTune framework is capable of identifying an optimal $s^*$ setting distinct from the default RHIC configuration ($s^*_{x,y} = 0\text{ m}, 0\text{ m}$), even if the resulting luminosity improvement is marginal.

\subsection{Section Summary}

These RHIC experiments demonstrate that the GPTune framework is fully capable of online optimization under dynamic machine conditions, using a custom normalized luminosity metric. The framework successfully identifies local maxima during the tuning process and qualitatively suggests its ability to locate $s^*$ values corresponding to a potentially higher luminosity.

However, the sPHENIX ZDC signal remained relatively insensitive to $s^*$ shifts under these experimental conditions. The ZDC integrates the full longitudinal collision profile over a significantly long RMS bunch length ($\sigma_s \approx 0.9-1.2$ m), while the original optimization was intended for the sPHENIX MVTX detector. The MVTX detector offers much higher sensitivity by restricting acceptance to a narrow $\pm 10$ cm longitudinal region, designed for the nominal $0.2$ m bunch length.

As a result, although the optimization converged successfully to a local optimum, the relatively flat ZDC response reduced the available gradient information, the achievable improvement is constrained not by the optimization algorithm itself, but by the choice of observable imposed by current hardware conditions.

Meanwhile, to address the non-stationary nature of the luminosity signal, future implementations of GPTune should explicitly account for time-dependent drift in the objective function or use physics-informed GP with the known luminosity formula.
\section{\label{sec: Discussion} Summary and Discussion}

In this paper, we have demonstrated the robust optimization capabilities of the GPTune framework across two distinct accelerator subsystems at Brookhaven National Laboratory. 

First, the framework was applied to EBIS intensity optimization, where it successfully achieved a $70\%$ intensity improvement at the XF14 CT. Notably, these gains were realized in a high-noise environment characterized by a $\pm 10\%$ standard deviation in the target signal (the device inherent characteristics), highlighting the algorithm's resilience to experimental fluctuations.

Second, following its successful application to light sources, FELs, and linacs within the accelerator community, this work extends the use of the BO algorithm to online collider luminosity optimization via $s^*$ control at RHIC. Despite the "flat" response of the ZDC caused by the $1.2 \text{ m}$ bunch length, GPTune demonstrated a clear ability to identify local maxima and successfully characterized residual $s_{x,y}^*$ offsets in the IR8 region.

Furthermore, the methodologies developed here, specifically the $s^*$ control method, are important for the upcoming Electron-Ion Collider (EIC) \cite{AbdulKhalek_Snowmass_2022}. Given the significantly shorter RMS bunch lengths projected for the EIC ($0.060\text{-}0.075\text{ m}$ for protons and $0.007\text{-}0.009\text{ m}$ for electrons), luminosity sensitivity to $s^*$ offsets will be much higher than in current operations. 

Therefore, the integration of the BO framework with $s^*$ control is now positioned as a important tool for the next generation of short bunch collider, such as the future Electron-Ion Collider (EIC).

\section*{Acknowledgments}

This work is supported by the U.S. Department of Energy (DOE) Office of Nuclear Physics under Award Number DE-SC0023518. This work is also supported by Brookhaven Science Associates, LLC under Contract No. DE-SC0012704, and by the DOE Office of Science, Office of Advanced Scientific Computing Research, Applied Mathematics program under Contract No. DE-AC02-05CH11231 at Lawrence Berkeley National Laboratory, and Contract No. DE-AC02-06CH11357.

The authors would also like to acknowledge A. Fedotov for his helpful suggestion.

\bibliographystyle{elsarticle-num} 
\bibliography{biblio.bib}

\end{document}